\def\expec#1{\langle#1\rangle}
\def\nothing{\noindent\centerline{\,}}

\def\etal{{\frenchspacing\it et al.}}
\def\ie{{\frenchspacing\it i.e.}}
\def\eg{{\frenchspacing\it e.g.}}
\def\etc{{\frenchspacing\it etc.}}
\def\rms{{\frenchspacing r.m.s.}}

\def\pp{\noindent\parshape 2 0truecm 13.6truecm 1truecm 12.6truecm}
\def\rf#1;#2;#3;#4 {\par\pp#1, {\it #2}, {\bf #3}, #4. \par}
\def\rg#1;#2;#3;#4;#5 {\par\pp#1, {\it #2}, {\bf #3}, #4 (``#5"). 
\par}
\def\rn{\pp}

\def\beq#1{\begin{equation}\label{#1}}
\def\eeq{\end{equation}}
\def\beqa#1{\begin{eqnarray}\label{#1}}
\def\eeqa{\end{eqnarray}}
\def\eq#1{equation~(\ref{#1})}

\def\bfig{\begin{figure}[h] \centerline{\hbox{}}\vfill}
\def\efig{\end{figure}\vfill\newpage}

\def\fheight{12cm}
\def\fwidth{17cm}
\def\fig#1{Figure~\ref{#1}}

\def\spose#1{\hbox to 0pt{#1\hss}}
\def\simlt{\mathrel{\spose{\lower 3pt\hbox{$\mathchar"218$}}
     \raise 2.0pt\hbox{$\mathchar"13C$}}}
\def\simgt{\mathrel{\spose{\lower 3pt\hbox{$\mathchar"218$}}
     \raise 2.0pt\hbox{$\mathchar"13E$}}}
\def\simpropto{\mathrel{\spose{\lower 3pt\hbox{$\mathchar"218$}}
     \raise 2.0pt\hbox{$\propto$}}}

\def\addr#1{{\small\it #1}}
\def\auth#1{{#1}}

\def\rh{\widehat{\bf r}}
\def\N{N}
\def\d{d}
\def\a{a}
\def\scale{\eta}
\def\l{\ell}
\def\Ylm{Y_{\l m}}
\def\wlm{w_{\l m}}
\def\summ{\sum_{m=-\l}^{\l}}

\def\ed{\end{document}}

\documentstyle[11pt,epsf,rotate]{article}
\begin{document}


\begin{titlepage}   

\noindent
\today
\hfill MPI-PhT/96-48
\begin{center}

\vskip0.9truecm
{\bf

AN ICOSAHEDRON-BASED METHOD FOR PIXELIZING THE CELESTIAL SPHERE\footnote{
Published in {\it ApJ Letters}, {\bf 470}, L81.\\
Both this paper and the source code available from\\
{\it h t t p://www.sns.ias.edu/$\tilde{~}$max/icosahedron.html} 
(faster from the US) and from\\
{\it h t t p://www.mpa-garching.mpg.de/$\tilde{~}$max/icosahedron.html} 
(faster from Europe).\\
Note that figures 2 and 3 will print in color if your printer supports it.
}
}

\vskip 0.5truecm
  \auth{Max Tegmark}
  \smallskip

  \addr{Max-Planck-Institut f\"ur Physik, F\"ohringer Ring 6,}
  \addr{D-80805 M\"unchen;}

 \addr{Max-Planck-Institut f\"ur Astrophysik,  
 Karl-Schwarzschild-Str. 1, D-85740 Garching;}\\

 \addr{email: max@ias.edu}

  \smallskip
  \vskip 0.2truecm

\end{center}

\begin{abstract}
\end{abstract}
\nothing\vskip-2.1truecm\nothing
\begin{center}
\makebox{
\baselineskip 0.4cm 
\parbox[l]{7.5truecm}{
For power spectrum estimation\\
it's important that the pixelization \\
of a CMB sky map be\\
smooth and regular to high degree.\\
With this criterion in mind\\
the ``COBE sky cube" was defined.\\
This paper has as central theme\\
to further improve on this elegant scheme\\
which uses a cube as projective base\\
- here an icosahedron is used in its place.\\
Although the sky cube is excellent,\\
a further reduction of 20 percent\\
}
\hglue0.3truecm
\parbox[r]{8.5truecm}{
of the number of pixels can be obtained\\
while the pixel distance is maintained,\\
and without any degradation\\
of accuracy for integration.\\
The pixels are rounder in this scheme where\\
they are hexagonal rather than square,\\
and the faces are small in this implementation\\
which simplifies area-equalization.\\
The reason distortion is lessened is that\\
the faces are smaller and therefore more flat.\\
To use the method, you can get\\
a FORTRAN code from the Internet.\\
}
}
\end{center}

\end{titlepage}


\section{INTRODUCTION}

In astronomy and cosmology, this is the age of map-making.
Recent ground-based and satellite-borne experiments have produced 
all-sky maps at a wide range of wavelengths, spanning from radio 
frequencies to the infra-red, ultraviolet and x-ray bands.
Although the brightness distributions being measured are always 
continuous functions of position, the maps are in practice compiled and 
distributed with values only at some finite number of points, or {\it 
pixels}. As discussed below, a good choice of pixelization scheme can 
often substantially simplify the subsequent analysis of the data, and 
for this reason, considerable amounts of work have been spent on 
developing good schemes for pixelizing the celestial sphere. Arguably 
the best and most elaborate method to date is the so-called {\it COBE 
sky cube} scheme (Chan \& O'Neill 1976, O'Neill \& Laubscher 1976),
which has been successfully employed for the 
DMR, DIRBE and FIRAS maps of the COBE satellite. This method has a 
number of desirable properties, and it is rather obvious that it
cannot be radically improved upon. 
However, the next generation of cosmic microwave background (CMB) maps 
from instruments such as the MAP and COBRAS/SAMBA satellites will be 
subjected to very extensive and time-consuming processing in order to 
obtain measurements of cosmological parameters. In view of this, it is 
timely to search for still better pixelization schemes, since
even quite modest improvements can translate into substantial 
numerical gains. The purpose of this {\it Letter} is to present such 
an improved method for pixelizing the sphere, akin to the sky cube 
method but replacing the cube by an icosahedron.

\subsection{What is a ``good" pixelization?}

What do we mean by a pixelization scheme being good? Specifically, if 
we are to place $\N$ points (pixel centers) on the sphere, where is 
the best place to put them? We will use the following two criteria:
\begin{enumerate}
\item The worst-case distance to the nearest pixel should be minimized.
\item We should be able to accurately approximate integrals by sums.
\end{enumerate}
Defining $d$ as the maximum distance that a point on the sphere can be 
from the pixel closest to it, criterion 1 says to minimize 
$\d$. 
Criterion 2 states that the integral of a function over the 
sphere should be well approximated by $(4\pi/N)$ times the sum of the 
function values at the pixel locations. This is important for 
applications such as CMB maps, where one wants to expand the 
brightness distribution in some set of functions, {\eg}, spherical 
harmonics.
Intuitively, we expect that both of these goals can be attained if the 
pixel distribution is in some sense as regular as possible. So if 
$N=6$, for instance, one might opt for the 6 corners of a regular 
octahedron. Unfortunately, there is only a finite number of platonic 
solids (SO(3) has only a finite number of discrete subgroups), so 
there is in general no obvious ``most regular" pixelization scheme.

\subsection{The icosahedron advantage}

\begin{figure}[phbt]
\hglue1.5truecm{{\vbox{\epsfxsize=18cm\epsfbox{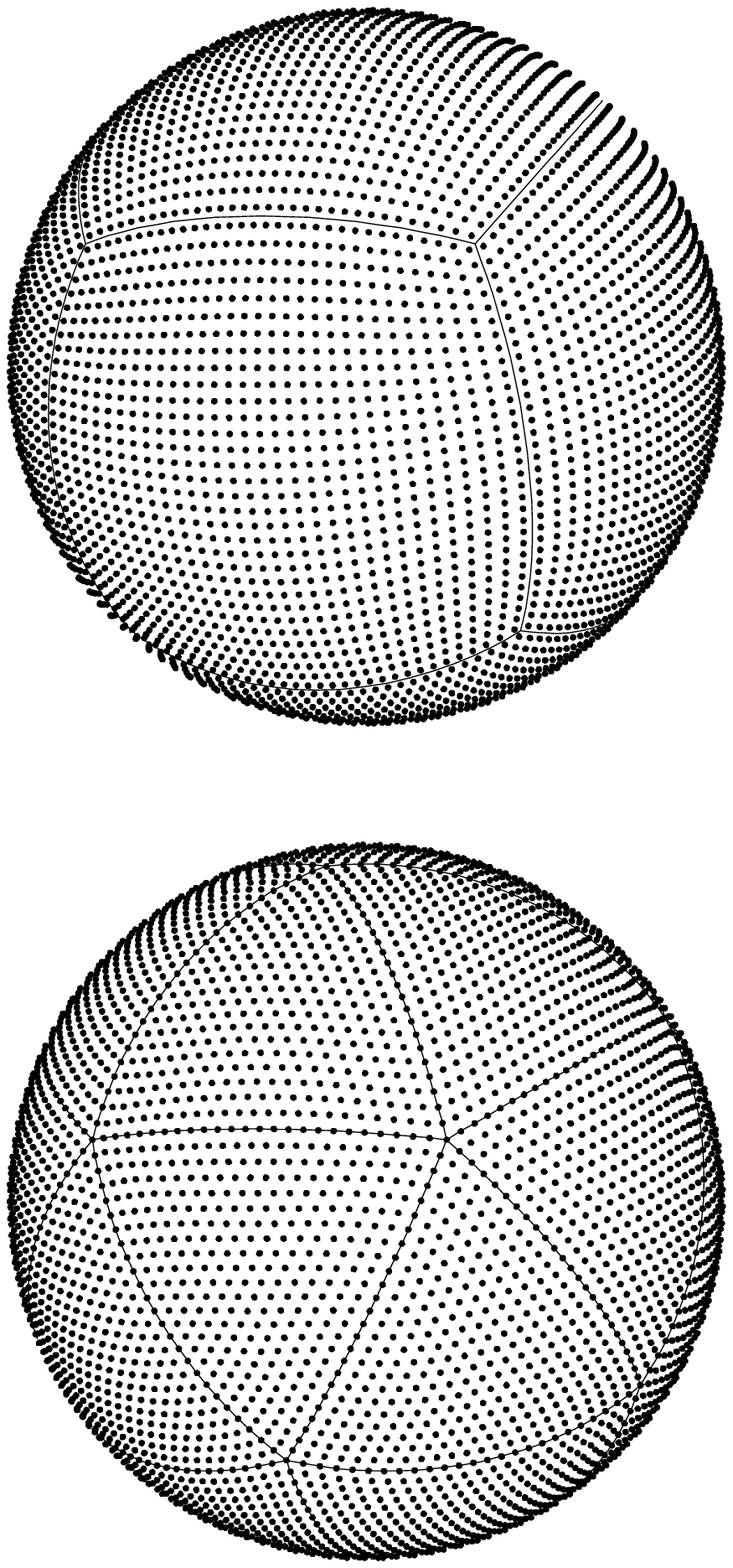}}}}
\caption{
The cube-based and icosahedron-based pixelization schemes.
}
\label{BallsFig}
\end{figure}

The COBE sky cube pixelization scheme is illustrated in \fig{BallsFig} (top), 
and consists of the following steps:
\begin{enumerate}
\item The sphere is inscribed in a cube, whose faces are pixelized 
with a regular square grid.
\item The points are mapped radially onto the sphere.
\item The points are shifted around slightly, to give all pixels 
approximately equal area.
\end{enumerate}
A pixel (the area which is closer to a given point than to all other 
points) is thus approximately square, 
with a side of length $\sim\sqrt{4\pi/N}$.
The points furthest from the pixels lie at the corners of these 
squares, so 
\beq{dCubeEq}
\d_{cube} \approx\sqrt{2\pi\over N}.
\eeq
For a honeycomb grid 
as illustrated in \fig{HexagonFig}, a pixel is hexagonal, and one 
readily computes that  
\beq{dIcosaEq}
\d_{icosa} \approx\sqrt{8\pi\over 3\sqrt{3}N},
\eeq
a value which is about $12\%$ smaller than that for the square grid 
case. To take advantage of this, one could thus replace the sky cube 
by a Platonic solid with triangular faces, {\ie}, by the tetrahedron, 
the octahedron or the icosahedron. 

The above-mentioned area-equalization is carried out for the sake of 
our second criterion, loosely speaking so that 
the equal weights that the pixels get when summed 
over correspond to equal weights $d\Omega$ in an integral 
(we present a rigorous application of criterion 2 in the discussion section).
Since the pixels originally were on a rectangular grid on the cube faces
(on the the
tangent plane of the sphere), the amount of ``stretching" required
increases toward the edges of the faces. Both this and the radial projection 
makes the pixels slightly deformed, so that the further out on a face one goes, 
the more the corresponding pixels on the sphere depart from a regular grid.  
Because of this, it is clearly desirable to use 
{\it as small faces as possible}, so that the corresponding 
regions of the sphere
are as flat as possible.
The Platonic solid with the smallest 
faces is the one with the largest number of faces: the icosahedron, 
whose faces are 20 triangles (see \fig{BallsFig}). Not only does it have 
the advantage of having more than three times as many faces as the cube, 
which one would expect to help with criterion 2, but 
since the faces are triangles rather than squares, 
it is better according to criterion 1 as well.
These advantages led Ned Wright to make an independent implementation
of the icosahedron scheme in the early 1980's, 
with an area-equalization that was approximate rather than exact. 

The rest of this {\it Letter} is organized as follows. 
The  icosahedron-based pixelization method is described in Section 2
and numerically compared with the COBE sky cube method in Section 3.

\section{METHOD}

The icosahedron pixelization scheme is illustrated in \fig{BallsFig} (bottom), 
and is akin in spirit to the COBE sky cube method:
\begin{enumerate}
\item The sphere is inscribed in an icosahedron, whose faces are pixelized 
with a regular triangular grid.
\item The points are mapped radially onto the sphere.
\item The points are shifted around slightly, to give all pixels 
approximately equal area.
\end{enumerate}
A FORTRAN package implementing this is available at \\
{\bf h t t p://www.sns.ias.edu/$\tilde{~}$max/icosahedron.html}\\ 
(faster from the US) and from\\
{\bf h t t p://www.mpa-garching.mpg.de/$\tilde{~}$max/icosahedron.html}\\ 
(faster from Europe).\\
The user interface is identical to that for the COBE sky cube package: 
for any specified resolution, one subroutine converts a pixel number to a
unit vector, and a second subroutine converts a unit vector to a pixel number,
the number of the pixel closest to that vector. Below we merely  
summarize the geometrical issues that specify the method.

\subsection{Part I: mapping to and from the icosahedron}

The 3D aspects of the problem are computationally trivial, 
since any of the 20 icosahedron faces can be rotated to lie in 
the $z=1$ plane by multiplication by an appropriate rotation matrix,
and all these rotation matrices can be precomputed once and for all.
The mapping between the $z=1$ tangent plane and the surface of the 
unit sphere preserves the direction of a vector and simply changes 
its length appropriately, either to be unity (on the sphere) 
or to have $z=1$. It is easy to see that straight 
lines on the tangent
plane correspond to great circles on the sphere. Thus each 
icosahedron face gets mapped onto a region on the sphere bounded
by three great circles. 

\subsection{Part II: the area equalization}

\begin{figure}[phbt]
\centerline{\epsfxsize=16cm\epsfysize=16cm\epsfbox{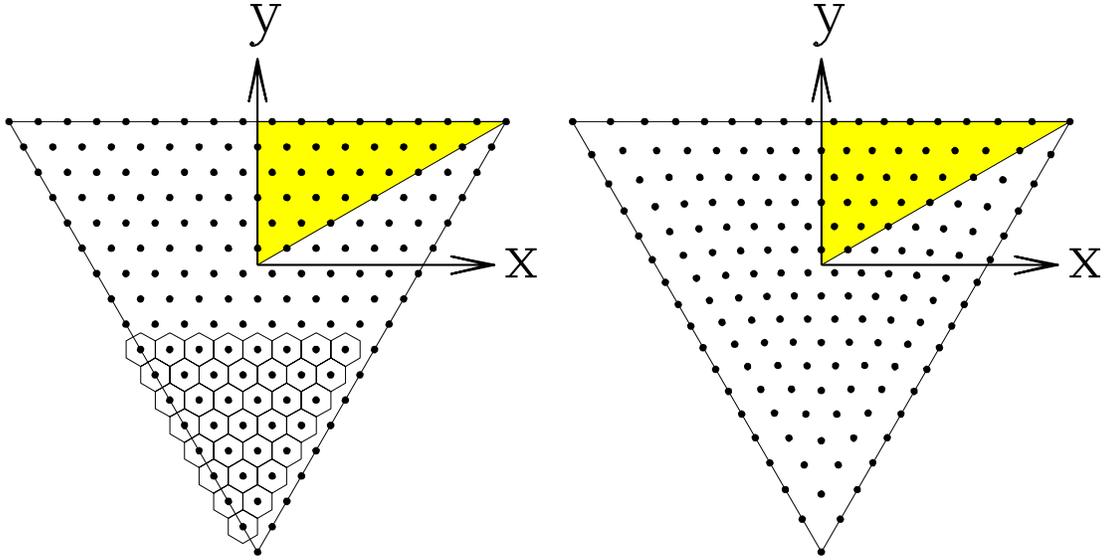}}
\vskip-6truecm
\caption{
A regular triangular grid (left) is adjusted (right)
to give all pixels the same area. As illustrated, the pixels 
have a hexagonal shape. A triangular icosahedron face can be symmetrically
decomposed into six identical right triangles (one is shaded), 
and the area equalization mapping is seen to respect this symmetry.
}
\label{HexagonFig}
\end{figure}

The area equalization step is illustrated in \fig{HexagonFig}.
After mapping part of the sphere onto a triangle in 
the tangent plane as above, 
we want to map this triangle onto itself
(``shift the pixels around") in such a way that the combined mapping 
becomes an equal-area mapping, {\ie}, gets a constant Jacobian.
The Jacobian of the mapping from the sphere to the plane is 
\beq{JacobianEq1}
\left|{\partial(x,y)/\partial\Omega}\right| = (1+x^2+y^2)^{3/2},
\eeq
so we want to find find a second mapping 
$(x,y)\mapsto (x',y')$ whose Jacobian is proportional to the inverse of this.
In other words, we wish to find two functions $(x',y')$ that
map the boundary of the triangle onto itself
and satisfy the nonlinear 
partial differential equation  
\beq{JacobianEq2}
\det \left(\matrix{
\partial x'/\partial x&\partial x'/\partial y\cr
\partial y'/\partial x&\partial y'/\partial y\cr}\right)
=\scale^2 (1+x^2+y^2)^{-3/2}
\eeq
for some proportionality constant $\scale$.
Since the icosahedron has 20 faces, the area of the triangular 
region on the sphere is clearly $4\pi/20$. 
The sides of the equilateral triangle in the tangent plane have length
$\a = \left[9\tan^2\left({\pi\over 5}\right) - 3\right]^{1/2}\approx 1.323$,
so its area is $\a^2 \sqrt{3}/4$.
Taking the ratio of these two areas fixes the above proportionality 
constant to be
\beq{ScaleEq}
\scale = 
\left[{15\sqrt{3}\over 4\pi}\left(3\tan^2{\pi\over 5}\,-\,1\right)\right]^{1/2}
\approx 1.098.
\eeq
The partial differential equation \eq{JacobianEq2} is under-determined
and admits infinitely many solutions, which allows us to impose 
additional simplifying requirements. As illustrated
in \fig{HexagonFig}, the triangle can be decomposed into six right triangles
of identical shape that can all be mapped into the one in
the upper right corner (shaded) by a combination of
$120^\circ$ rotations and reflections. We require our solution to 
respect this symmetry, so we merely need to find a solution to 
\eq{JacobianEq2} in the shaded triangle that maps its 
boundary onto itself. We use the additional freedom to require
that horizontal lines in this region get mapped onto horizontal lines.
This is enough to determine the solution uniquely, and we find that
\beq{ForwardMapEq}
\cases{
y'&$=
\scale\sqrt{
{2\over\sqrt{3}}
\tan^{-1}\left[\sqrt{3}
{\sqrt{1+4y^2}-1\over\sqrt{1+4y^2} + 3}
\right]}$,\cr
\noalign{\vskip 4pt}
x'&$=\left({\scale xy'\over y}\right)
\sqrt{1+4y^2\over 1+x^2+y^2}$,
}
\eeq
which can be verified by direct substitution.
These equations are readily inverted, giving 
\beq{BackwardMapEq}
\cases{
y&$=
{1\over 2}\sqrt{
3\left[
{1+\sqrt{3}\tan(\sqrt{3}y'^2/2\scale^2)\over
\sqrt{3}-\tan(\sqrt{3}y'^2/2\scale^2)}
\right]^2 - 1}
$,\cr
\noalign{\vskip 4pt}
x&$=x'y\sqrt{{1+y^2\over y'^2(1+4 y^2)-x'^2 y^2}}$.
}
\eeq
This area-equalizing mapping is 
illustrated in \fig{HexagonFig}, where the regular triangular 
grid of points (left) has been adjusted (right) to give equal-area
pixels when projected onto the sphere. The pixels in \fig{BallsFig}
have also been equal-area adjusted --- otherwise a slight excess would 
be visible near the corners of the triangles.

\section{DISCUSSION}

We have presented a new method for pixelizing the sphere, devised
to be useful for storing and analyzing all-sky maps
in astronomy and cosmology, and made a FORTRAN 
implementation publicly available over the Internet.

As far as practical issues goes, it is essentially 
equivalent to the COBE sky cube method: 
the pair of subroutines that convert between 
unit vectors and pixel numbers are for all practical purposes
instantaneous. How does its geometric performance compare with 
that of the COBE sky cube method according to the two criteria 
described in the introduction?
As discused, the fact that the pixels are hexagons rather than 
squares reduces the maximum
distance to the grid by about $10\%$.\footnote{
Other natural benchmarks such as the average and {\rms} distances 
to the grid get reduced by a similar factor.
}
With respect to criterion 1, 
it is easy to see that hexagonal pixels are optimal 
on a flat surface, so they clearly cannot be 
substantially improved upon for the sphere either when 
$N$ is large. In addition, rounder pixels are of course
appealing since the instrumental beam tends to be round.
We will now examine criterion 2 in more detail, and find that in 
a certain well-defined sense, the improvement of the icosahedron method 
when integrating is about $10\%$ as well.

\subsection{Spherical Cubature}

The study of how to best approximate integrals with sums
has a long tradition in the mathematics literature.
For instance, the famous quadrature formula of Gauss shows
how a 1-dimensional integral $\int_a^b f(x)dx$ can be approximated 
with a weighted average $\sum_{i=1}^N w_i f(x_i)$
such that the approximation becomes exact if 
$f$ is a polynomial of degree less than $2N$. 
That this is plausible can be readily seen by noting that 
there are $2N$ free parameters (the positions $x_i$ and the
weights $w_i$) available to satisfy the $2N$ constraints.  
When integrating on a circle rather than an interval, the Gauss
problem becomes greatly simplified, and a simple Fourier expansion
shows that exactness for polynomials of degree less than $2N$ is obtained
by the most naive prescription possible: 
$N$ equispaced points with equal weights (as compared with the
zeroes of the Legendre polynomials in the Gauss case).
In other words, the 1-D interval case appears to have been complicated
by the presense of endpoints, whereas in the fully symmetric case, 
the optimal scheme was that where the pixelization was as 
regular as possible. Since the sphere also lacks endpoints that
break symmetry, one might therefore conjecture that the optimal 
integration formula would involve a maximally regular pixelization 
and equal weights. Unfortunately, it is a well-known group-theoretical result
that there are no completely regular point distributions on the sphere
for $N>20$. This has led to an extensive body of work on the
problem of optimal integration on the sphere --- see {\eg}
Stroud (1971), Sobolev (1974), Konjaev (1979) and Mysovskikh (1980)
for theoretical work on this so-called cubature problem. 
Although no strictly optimal method has been found for 
general $N$ (which is one of the main foci of the mathematics literature, 
together with proofs of various bounds on how well one can do), 
we will see that from a pragmatic astrophysicist's point of view, 
the icosahedron scheme is so close to optimal that further improvements 
may not be worthwhile. 

\begin{figure}[phbt]
\centerline{{\vbox{\epsfxsize=16cm\epsfysize=12cm\epsfbox{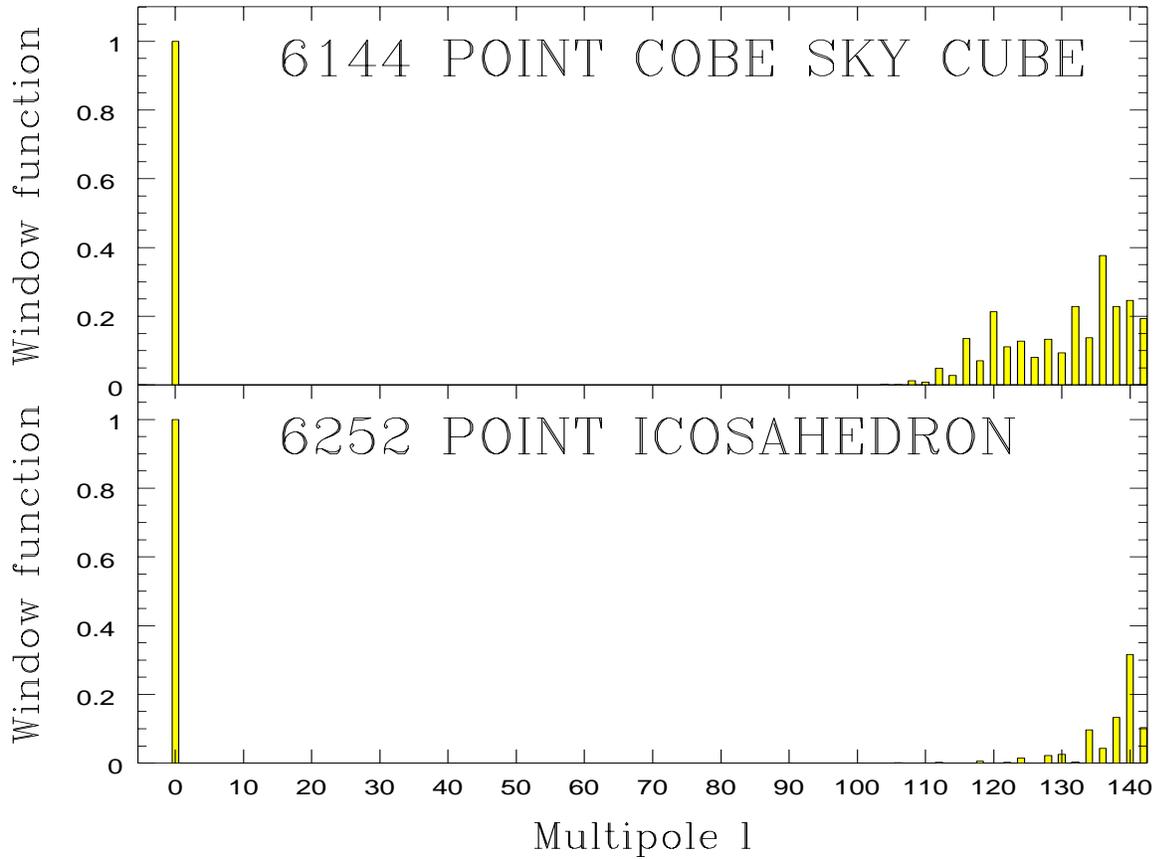}}
}}
\caption{
Window function comparison. The histograms show the 
errors obtained when approximating integrals by sums, multipole 
by multipole. The icosahedron method generally remains accurate down
to about $10\%$ smaller scales than a COBE sky cube with a similar 
number of pixels, which means that it can produce comparable 
results using about $20\%$ fewer pixels.
}
\label{WindowFig}
\end{figure}

We can clearly write our approximation of the integral 
$\int f d\Omega$ as $\int w f d\Omega$, where the weight function $w(\rh)$ is a
linear combination of $N$ delta-functions. 
Let us define the {\it integration error} as 
\beq{DeltaDefEq}
\Delta\equiv \sum_{i=1}^N w_i f(\rh_i) - \int f(\rh)d\Omega
= \int [w(\rh)-1] f(\rh) d\Omega.
\eeq
The cubature problem thus involves finding a $w$ that makes $\Delta$ 
vanish when $f$ is any polynomial up to a given degree.
We see that this is equivalent to finding a $w$ that is orthogonal to 
all such polynomials except the monopole (which gives the integral).
As an orthonormal basis for distributions on the sphere,
let us select the Gram-Schmidt orthogonalized polynomials.
This basis is
simply the spherical harmonics $\Ylm$, where $\l$ gives the degree of
the polynomials ($\l=1$ gives linear functions, $\l=2$ 
gives harmonic quadratic polynomials, {\etc}). A 
useful way to diagnose any integration scheme is thus to compute the
spherical harmonic coefficients of $w$,
\beq{flmDefEq}
\wlm \equiv \int\Ylm^*(\rh)w(\rh)d\Omega = 
\sum_{i=1}^N w_i \Ylm^*(\rh_i),
\eeq
and plot its {\it window function} $W_\l$, defined as
\beq{WdefEq}
W_\l \equiv\summ|\wlm|^2.
\eeq
Such histograms are plotted in \fig{WindowFig} for the COBE sky cube method
and the icosahedron method using a comparable number of points,
both with all weights $w_i=4\pi/N$.
Apart from the monopole (which gives the integral), both are seen to
vanish for all polynomials of degree $\l\ll 100$, which means that these
methods are essentially exact in Gauss' sense to that order.
Comparing the two methods, the icosahedron scheme is seen to remain 
accurate out to approximately $10\%$ greater $\l$-values, so in this 
sense, the method is about $10\%$ better. This gain factor was found
be remain around $10\%$ over the range of $N$-values likely to be of 
astrophysical interest.

How close to optimal is the icosahedron method?
Although a rigorous lower bound for general $N$ has still not been 
proven, an approximate answer can readily be found by simple 
constraint-counting as with the Gaussian quadrature case above.
There are $\l^2$ spherical harmonics of degree less than $\l$,
whereas $w$ is specified by $3N$ free parameters 
($N$ weights $w_i$ and $N$ unit vectors $\rh_i$).
One might thus hope to obtain a perfect window function
up to $\l\approx\sqrt{3N}$, {\ie}, to make the integration exact
for polynomials of degree up to $\sqrt{3N}$.
For the examples in \fig{WindowFig}, we have
$\sqrt{3N}\approx 136$ and $137$, respectively, {\ie}, 
values quite close to where the icosahedron window function becomes
substantial. 
The problem of finding a strictly optimal solution has
been attacked numerically (Schmid 1978), but the nonlinear system
of equations involved was found very difficult to solve in 
practice for large $N$. Moreover, from the point of view of a 
pragmatic astrophysicist,
it is not even clear that such a solution would be better than those in
\fig{WindowFig}, since there is no guarantee that its window function
does not explode uncontrollably for $\l>\sqrt{3N}$. 
If the mapped signal has some angular power spectrum $C_\l$, then
the mean square integration error is readily seen to be
\beq{IntErrEq}
\expec{\Delta^2} = \sum_{\l=1}^\infty W_\l C_\l.
\eeq
In astrophysics applications, the signal power spectrum typically 
falls off smoothly around the scale set by the beam width of the
observing instrument, so the mathematical problem of making
$W_\l$ {\it exactly} zero while ignoring $W_{\l+1}$ altogether is
clearly not physically motivated. Rather, the astrophysicists concern 
is simply that $W_\l\approx 0$ well beyond the beam smoothing scale, 
and then grows in a controlled way.

Finally, it should be emphasized that these integration-related issues
are crucial for the next generation of cosmic microwave background maps,
since they will be integrated against a large number of weight
functions in order to obtain accurately power spectrum
measurements. Since the data processing in these applications  
involves matrix algebra where the computational cost scales
as $N^3$ (Bunn \& Sugiyama 1995; Tegmark {\etal} 1996), 
even modest reductions in the number of pixels translate
into substantial savings in CPU time and storage requirements. 
For instance, a $12\%$ better window function allows $25\%$
fewer pixels, which corresponds to halving the CPU time.

Given the great efforts that will be spent on collecting and 
analyzing such data sets, there should be no reason to 
use anything but the best scheme when pixelizing the data.
We have found that the icosahedron method improves upon the
COBE sky cube method by about $10\%$ when it comes to 
both integration accuracy and worst-case
distance to the nearest pixel center. 
Since this
improvement is computationally speaking free, 
it is hoped that the icosahedron method will be useful for future
mapping experiments.

\bigskip
The author wishes to thank James Binney,
Ang\'elica de Oliveira-Costa, Vikram Seth, 
Harold Shapiro and Ned Wright 
for useful comments,
and Schwabinger Krankenhaus for hospitality
during the visit where this work was carried out.
This work was partially supported by European Union contract
CHRX-CT93-0120 and Deutsche Forschungsgemeinschaft grant SFB-375. 


\section{REFERENCES}

\rf Bunn, E. F. \& Sugiyama N. 1995;ApJ;446;49

\rn Chan, F.K. \& O'Neill, E.M. 1976,
{\it Feasibility study of a quadrilateralized spherical cube Earth data base},
Computer Sciences Corp. EPRF Technical Report.

\rf Konjaev, S. I. 1979;Mat. Zametki;25;629

\rn Mysovskikh, I. P. 1976, in {\it Quantitative Approximation},
eds. R. A. Devore \& K. Scherer (New York: Academic Press).

\rn O'Neill, E.M. \& Laubscher, R.E. 1976,
{\it Extended studies of a quadrilateralized spherical
cube Earth data base}, Computer Sciences Corp. EPRF Technical Report.

\rf Schmid, H. J. 1978;Numer. Math.;31;281

\rn Sobolev, S. L. 1974, 
{\it Introduction to the theory of cubature formulae} 
(Moscow: NAUKA).

\rn Stroud, A. H. 1971, 
{\it Approximate Calculation of Multiple Integrals} 
(Englewood Cliffs: Prentice-Hall).

\rn Tegmark, M., Taylor, A. \& Heavens, A. F. 1996, 
preprint astro-ph/9603021.

 
\def\fheight{10.3cm} \def\fwidth{14.5cm}

\clearpage

\clearpage

\clearpage

\end{document}